\documentclass[aps,showpacs,nofootinbib,twocolumn]{revtex4}

\usepackage{latexsym}
\usepackage{epsfig}
\usepackage{amsmath}
\usepackage{amssymb}
\usepackage{graphicx}

\newcommand{\lp}{\left(}
\newcommand{\rp}{\right)}

\newcommand{\ba}{\begin{eqnarray}}
\newcommand{\ea}{\end{eqnarray}}
\newcommand{\be}{\begin{equation}}
\newcommand{\ee}{\end{equation}}

\newcommand{\beq}{\begin{equation}}
\newcommand{\eeq}{\end{equation}}
\newcommand{\bea}{\begin{eqnarray}}
\newcommand{\eea}{\end{eqnarray}}
\newcommand{\bseq}{\begin{subequations}}
\newcommand{\eseq}{\end{subequations}}

\newcommand{\Ref}[1]{(\ref{#1})}

\usepackage{color}

\begin{document}

\title{Einstein static Universe in non-minimal kinetic coupled gravity}

\author{K. Atazadeh}\email{atazadeh@azaruniv.ac.ir}
\author{F. Darabi}\email{f.darabi@azaruniv.ac.ir}

\affiliation {Department of Physics, Azarbaijan Shahid Madani University, Tabriz, 53714-161 Iran}

\date{\today}

\begin{abstract}

 We study the stability of Einstein static Universe, with FLRW metric, by considering linear homogeneous perturbations in the kinetic coupled gravity. By taking linear homogeneous perturbations, we find that the stability of Einstein static Universe, in the kinetic coupled gravity with quadratic scalar field potential, for closed ($K=1$) isotropic and homogeneous FLRW Universe depends on the coupling parameters $\kappa$ and $\varepsilon$. Specifically, for $\kappa=L_P^2$ and $\varepsilon=1$ we find that the stability condition imposes the inequality $a_0>\sqrt{3}L_P$ on the initial size $a_0$ of the closed Einstein static Universe before the inflation. Such inequality asserts that the initial size of the Einstein static Universe must be greater than the Planck length $L_P$, in consistency with the quantum gravity and quantum cosmology requirements. In this way, we have determined the non-minimal coupling parameter $\kappa$ in the context of Einstein static Universe. Such a very small parameter is favored in the inflationary models constructed in the kinetic coupled gravity. We have also studied the stability against the vector and tensor perturbations and discussed on the acceptable values of the equation of state parameter.

\end{abstract}

\pacs{04.20.Jb, 04.50.Kd, 95.36.+x}

\maketitle

\section{Introduction}
Inflationary scenario can address most of the problems in the standard cosmology, however, in spite of the interesting prosperities of inflationary scenario,
the existence of a big bang singularity at  the beginning of  Universe
is the major problem of standard cosmology. In attempt to remove the initial singularity,
several theories  have been proposed to address this issue, such as the string/M-theory, the pre-big bang theory \cite{1} and ekpyrotic/cyclic
 \cite{2}.

In the static closed Friedmann-Lema\^{\i}tre-Robertson-Walker model, the Einstein static Universe is one of the exact solutions of Einstein's equations coupled to  a perfect fluid and a cosmological constant \cite{3}. The stability conditions of Einstein static Universe have been widely studied in the literature indicating that this solution is not usually stable against the homogeneous perturbations \cite{4}. In addition, it has been shown that this solution has neutral stability against the adiabatic scalar inhomogeneities with high
enough sound speed, as well as the small inhomogeneous vector and tensor perturbations \cite{5}.
Nevertheless, it was  shown that the Einstein static Universe is unstable against Bianchi type-IX
spatially homogeneous perturbations in the presence of perfect fluids with $\rho + 3P > 0$ \cite{6},
and for various sources of matter fields  \cite{7}.
Regardless of historical importance of the Einstein static Universe, the reiterated interest to this solution comes from the ``Emergent Universe''
scenario, an inflationary cosmological model in which Einstein static Universe plays an incisive role as a initial state.

In the context of general relativity, this model was proposed in  2004 by Ellis {\it et al}  to solve the problem of initial singularity in the standard cosmological model  \cite{8}.
Moreover, the Einstein static Universe has been discussed in several modified gravitational theories and quantum gravity models.
Actually, when we are working with the modified cosmological equations, it is possible to find many new static solutions, essentially  different from that of  classical Einstein static solution of GR, in which the stability properties depend  on the details of the studied theory or family of theories taken into account. Basically, due to  the existence of neutral stable solutions,   the fine-tuning problem of cosmological constant is so improved. But, in fact a mechanism
is needed to finish  the phase of infinite expansions and collapses, and to operate the expanding phase of the Universe \cite{9}.
Such a mechanism has been known as ``inflation'' \cite{inflation}.

In the context of inflationary cosmology, the role of scalar field potential
to establish an inflation is unavoidable. In general, the slowly varying potentials behave like a large effective cosmological constant suitable for driving an inflation. The question that `` Is it possible to recover the cosmological constant and the inflationary phase "without" considering any effective potential'' led some authors to try for constructing an effective cosmological constant starting from extended gravity theories such as non-minimally coupled or higher order theories \cite{capo1}. In \cite{Amend}, the author considered some types of coupling between curvature and the scalar field, called non-minimal derivative coupling. The authors in \cite{capo2} studied this kind of couplings and connected them with inflation. Realistic cosmological scenario was introduced based
on non-minimal kinetic coupling and it was shown that at early Universe the domination of coupling term in the field equation predicts a quasi-de Sitter expansion \cite{10}. In the background of cosmological scenarios, the non-minimal kinetic coupling gravity has been considered by choosing zero and constant
potentials, for the quintessence and the phantom cases \cite{11}. Also, in \cite{11,111} the authors have considered some cosmological aspects of the non-minimal kinetic coupling gravity  such as Big Bang, an expanding Universe with no beginning, an eternally contracting Universe, a Big Crunch, a Big Rip avoidance and a cosmological bounce in the absence of the matter. In general, the scalar tensor theory of gravity with scalar field non-minimally coupled to gravity
reveals interesting cosmological and astrophysical behaviors \cite{Gao}, \cite{capoz}.

 According to the above approach to the issue of inflation, it is interesting to study the Einstein static Universe in the inflationary Universe based on the non-minimal kinetic coupled gravity. We show that an asymptotically Einstein static Universe in such inflationary Universe may result due to the terms in the field equations  of the non-minimal kinetic coupled gravity. In fact, we find that at early Universe these terms could be dominating and the cosmological evolution could have started around an Einstein static Universe with a size $a_0>\sqrt{\frac{3\kappa}{\varepsilon}}$, where $\kappa$ and $\varepsilon$ are coupling parameters (see below). We try to remove the initial singularity problem in the standard cosmological model by studying Einstein static Universe
  and its stability in the non-minimal kinetic coupled gravity theory. Actually, the stability of Einstein static state has been studied in various theories: in GR with a non-constant pressure \cite{ESa}, in brane world scenarios~\cite{Gergely:2001tn}, in Einstein-Cartan gravity~\cite{Boehmer:2003iv}, in loop quantum cosmology~\cite{Mulryne:2005ef}, in $f(R)$ gravity~\cite{Boehmer:2007tr,Goswami:2008fs,Seahra:2009ft}, in Gauss-Bonnet gravity~\cite{Bohmer:2009fc}, in IR modified Ho\v{r}ava gravity~\cite{Boehmer:2009yz}, in massive gravity~\cite{Parisi:2012cg}, and induced matter Brane Gravity \cite{Induced}.

This paper is organized as follows. In section~\ref{II}, we briefly review the formalism of the kinetic coupled gravity theory, in particular the action and field equations. In section~\ref{sect:3}, we present the  modified Friedman equations within the kinetic coupled gravity. In section~\ref{sect:4}, we consider linear homogeneous perturbations and study the stability of  Einstein static Universe in the kinetic coupled gravity. In section~\ref{sec:5}, we study the stability against the vector and tensor perturbations. We summarize our results in section~\ref{sec:concl}.
\section{Non-minimal kinetic coupling gravity\label{II}}

Let us consider a gravitational theory with non-minimal derivative coupling given by the action \cite{22}
\begin{eqnarray}\label{action}
S\hspace{-2mm}&=&\hspace{-3mm}\int d^4x\sqrt{-g}\left\{ \frac{R}{8\pi} -\hspace{-1mm}\big[\varepsilon
g_{\mu\nu} + \kappa G_{\mu\nu} \big] \phi^{,\mu}\phi^{,\nu} -2
V(\phi)\right\}\\\nonumber&&+S_m,
\end{eqnarray}
where ${S}_m$ stands for the action of matter,
$V(\phi)$ is a scalar field potential, $G_{\mu\nu}$ is the Einstein tensor, $\varepsilon$ takes the value +1 for the canonical field and -1 for the phantom one and $\kappa$ is the
coupling parameter with dimension of ({\em length})$^2$.
Varying the action (\ref{action}) with respect to $g_{\mu\nu}$ and $\phi$ gives the field equations, respectively:
\bseq\label{fieldeq}
\bea
\label{eineq}
&& G_{\mu\nu}=8\pi\big[T_{\mu\nu}^{(m)}+T_{\mu\nu}^{(\phi)}
+\kappa \Theta_{\mu\nu}\big], \\
\label{eqmo}
&&[\varepsilon g^{\mu\nu}+\kappa G^{\mu\nu}]\nabla_{\mu}\nabla_\nu\phi=V'(\phi),
\eea
\eseq
where $V'(\phi)\equiv dV(\phi)/d\phi$, $T^{(m)}_{\mu\nu}$ is a stress-energy
tensor of ordinary matter, and
\bea \label{T}
T^{(\phi)}_{\mu\nu}&=&\varepsilon[\nabla_\mu\phi\nabla_\nu\phi-
{\textstyle\frac12}g_{\mu\nu}(\nabla\phi)^2]-g_{\mu\nu}V(\phi), \\
\Theta_{\mu\nu}&=&-{\textstyle\frac12}\nabla_\mu\phi\,\nabla_\nu\phi\,R
+2\nabla_\alpha\phi\,\nabla_{(\mu}\phi R^\alpha_{\nu)}
\nonumber\\
&&
+\nabla^\alpha\phi\,\nabla^\beta\phi\,R_{\mu\alpha\nu\beta}
+\nabla_\mu\nabla^\alpha\phi\,\nabla_\nu\nabla_\alpha\phi
\nonumber\\
&&
-\nabla_\mu\nabla_\nu\phi\,\square\phi-{\textstyle\frac12}(\nabla\phi)^2
G_{\mu\nu}
\label{Theta}\\
&&
+g_{\mu\nu}\big[-{\textstyle\frac12}\nabla^\alpha\nabla^\beta\phi\,
\nabla_\alpha\nabla_\beta\phi
+{\textstyle\frac12}(\square\phi)^2
\nonumber\\
&& \ \ \ \ \ \ \ \ \ \ \ \ \ \ \ \ \ \  \ \ \   \ \ \  \ \ \ \ \ \
\ \ \ \ \ -\nabla_\alpha\phi\,\nabla_\beta\phi\,R^{\alpha\beta}
\big]. \nonumber
\eea
By imposing the Bianchi identity $\nabla^\mu G_{\mu\nu}=0$ and the matter conservation law
$\nabla^\mu T^{(m)}_{\mu\nu}=0$, equation \Ref{eineq} reduces to
\beq
\label{BianchiT}
\nabla^\mu\big[T^{(\phi)}_{\mu\nu}+\kappa \Theta_{\mu\nu}\big]=0.
\eeq
Note that by inserting equations \Ref{T} and \Ref{Theta} into \Ref{BianchiT} the differential equation \Ref{BianchiT} reduces to \Ref{eqmo}.
Simply, equation \Ref{eqmo} is a differential consequence of equation \Ref{eineq}.

The authors in \cite{111}, have established an inflation model without scalar field potential for the kinetic coupled gravity with spatially flat ($K=0$) FLRW metric and a cosmological constant, where the cosmological evolution of Universe at the vacuum dominated state $p_v=-\rho_v$ is described by
\begin{equation}\label{H_{kappa}}
a(t)\propto \exp(H_{\kappa}t),
\end{equation}
and
\begin{equation}\label{3H_{kappa}}
\dot{\phi}(t)\propto \exp(-3H_{\kappa}t),
\end{equation}
where
\begin{eqnarray}\label{Einstein1}
H\simeq\sqrt{\frac{1}{9\kappa}},
\end{eqnarray}
\begin{eqnarray}\label{Einstein2}
\dot{H}\simeq 0.
\end{eqnarray}
As is seen in (\ref{Einstein1}), the role of coupling parameter $\kappa$ in this inflationary behavior is important such that small value of $\kappa$ results in a sufficiently large value of the Hubble parameter. Although the present model is different from \cite{111}, regarding the scalar field potential and the curvature parameter $K$, but it is interesting to study the Einstein static Universe and its stability in the context of kinetic coupled gravity and investigate the possible impact of stability requirement on the coupling parameter $\kappa$.

\section{Einstein static Universe and modified Friedmann equations}\label{sect:3}

\subsection{Effective Friedmann equations}

The cosmological studies of the non-minimal kinetic coupling gravity have been sufficiently investigated \cite{10, 11, 111}. Especially, it was shown that the inflation and any cosmological behaviour, explicitly  depends on the non-minimally kinetic term of a scalar field $\phi$ with the curvature. However, it is  important to notice that the non-minimally kinetic term of a scalar field $\phi$ with the curvature describes further degrees of freedom of the gravitational field resulting from modified gravities.

We apply the Friedmann-Lema\^{\i}tre-Robertson-Walker (FLRW) line element as follows
\ba
ds^2=-dt^2+a^2(t)\left[ \frac{dr^2}{1-Kr^2} +r^2 (d^2 \theta + \sin^2\theta d^2\phi) \right],
\ea
where $K = +1, 0, -1$ denotes a closed, flat, and open Universe, respectively.

By including the effective energy density and pressure, the modified Friedmann equations can be written as \ba
3H^2 & = & 8\pi\rho_{\rm eff} - \frac{3K}{a^2}\,,
\label{fr1} \\
\dot{H} & = & -4\pi\lp\rho_{\rm eff}+p_{\rm eff}\rp + \frac{K}{a^2}
\label{fr2} \,,
\ea
where $\rho_{\rm eff}$ and $p_{\rm eff}$ are given by
\ba
\rho_{\rm eff} & = & \frac{1}{2}\dot{\phi}^2\Big[\varepsilon-3\kappa\Big(3H^{2}+\frac{K}{a^{2}}\Big)\Big] + V(\phi) +\rho_m\,,
\label{m1}\\
p_{\rm eff} & = & \frac{1}{2}\dot{\phi}^2\Big[\varepsilon+\kappa\Big(2\dot{H}+3H^{2}+\frac{K}{a^{2}}+4H\ddot{\phi}\dot{\phi}^{-1}\Big)\Big] \nonumber\\&&
-V(\phi) + p_m\,.
  \nonumber \\
\label{m2}
\ea

The conservation equations for the matter component and the scalar field are
given by\ba
\dot{\rho}_m + 3H(\rho_m + p_m) & = & 0\,,
\label{ma}\\
\varepsilon(\ddot{\phi} + 3H\dot{\phi})\hspace{-1mm}-\hspace{-1mm}3\kappa\Big[\Big(H^{2}+\frac{K}{a^{2}}\Big)\ddot{\phi}&&\nonumber\\+2H\dot{H}\dot{\phi}+3H^{3}\dot{\phi}+
\hspace{-1mm}\frac{KH\dot{\phi}}{a^{2}}\Big]\hspace{-2mm}&=-&\hspace{-2mm} V'(\phi)
\label{kg}\,.
\ea

\subsection{Einstein static Universe}

To study the  Einstein static Universe we impose $a = a_0 = {\rm const}$, thus $H=\dot{H}=0$ and also we take the matter distribution that obeys from the linear equation of state $p_m = w \rho_m$.  As a result, the effective Friedmann equations (\ref{fr1}) and~(\ref{fr2}) can be written as
\begin{align}
  8\pi\rho_{\rm eff} &= \frac{3K}{a_0^2}\,,
  \label{bg1}\\
  4\pi\lp\rho_{\rm eff}+p_{\rm eff}\rp &= \frac{K}{a_0^2}\,,
  \label{bg2}
\end{align}
respectively, implying the following condition which is imposed on the distribution of effective matter
\begin{align}
  \rho_{\rm eff} + 3 p_{\rm eff} = 0\,.
  \label{bg3}
\end{align}

By assuming $\phi = \phi_0 = {\rm const}$, the effective matter condition (\ref{bg3}) will imply
\begin{align}
  \frac{1}{2}\rho_m^{(0)}(1+3w) =  V(\phi_0) \,,
  \label{bg4}
\end{align}
\begin{align}
  \frac{K}{8\pi a_0^2}= \frac{V(\phi_0)(1+w)}{(1+3w)} \,,
  \label{bg44}
\end{align}
and finally for the modified Klein-Gordon equation (\ref{kg}) we will have
\begin{align}
  V'(\phi_0)=0, \,\,~~~\Lambda=8\pi V(\phi_0).
  \label{bg5}
\end{align}
Additionally, we can obtain  $a_0$  and $\rho_m$ in terms of $\phi_0$ and  $V(\phi_0)$.

\section{Stability analysis of the Einstein static Universe}\label{sect:4}

 The Einstein static Universe has been renewed as the asymptotic inspiration of an emergent Universe, to remove the initial singularity problem in the inflationary cosmology \cite{8}. Actually, these cosmological models contain remarkable features such as the absence of an initial singularity and avoidance of the quantum gravity intricacy. In a series of works by the present authors the stability analysis of the Einstein static Universe has been studied in different models. In \cite{Boehmer:2003iv}, the existence and stability of the Einstein static Universe have been studied in the Einstein-Cartan gravity and shown that this Universe in the presence of perfect fluid with spin density satisfying the Weyssenhoff restriction is cyclically stable around a center equilibrium point. In \cite{Mulryne:2005ef}, the stability of Einstein static Universe against the homogeneous scalar perturbations in the context of braneworld scenario is investigated and the stability regions are obtained in terms of the constant geometric linear equation of state parameter for the case of closed, open or flat Universe. It is also found that a stable Einstein static Universe may exist in a braneworld theory of gravity against scalar, vector and tensor perturbations for some suitable values and ranges of the cosmological parameters. In \cite{Boehmer:2009yz}, the stability of Einstein static Universe versus the linear scalar, vector and tensor perturbations is investigated in the context of deformed Ho\v{r}ava-Lifshitz (HL) cosmology inspired by entropic force scenario. It is shown that there is no stable Einstein static Universe for the case of flat Universe, however, for the closed Universe and large values of running parameter of HL gravity there is stability with domination of the quintessence and phantom matter fields, and for open Universe there is stability with domination of the matter fields. A neutral stability against the vector perturbations is obtained and it is shown that for large values of the running parameter of HL gravity, there is a stability against the tensor perturbations. In \cite{Induced}, the stability of Einstein static Universe against the scalar, vector and tensor perturbations
in the context of induced matter brane gravity is investigated. It is found that a stable Einstein static Universe against the scalar perturbations does exist provided that the variation of time dependent geometrical equation of state parameter is proportional to the minus of the variation of the scale factor. In all these works, we were motivated to find any impact of stability requirement of Einstein static Universe on the physical characteristics of modified theories of gravity.

Following the above mentioned line of investigation in the context of modified theories of gravity, in the present work, first we aim to study the stability analysis of Einstein static Universe against the homogeneous perturbations in the kinetic coupled theory of gravity. Our motivation is to find the possible impact of imposing such stability requirement on the kinetic coupling features of kinetic coupled theory of gravity.

To consider the stability of the Einstein static Universe in the present model, we take the following homogeneous perturbations
\begin{align}
  a &= a_0 + \delta a(t)\,,\\
  \phi &= \phi_0 + \delta \phi(t)\,,\\
  \rho_m &= \rho^{(0)}_m + \delta \rho_m(t).
\end{align}
Here, we only discuss on the adiabatic perturbations $\delta p_m(t) = w\, \delta \rho_m(t)$. By perturbing the field equations~(\ref{fr1}) and~(\ref{fr2}), and the evolution equations for the matter~(\ref{ma}) and the scalar field ~(\ref{kg}), we only study perturbations up to linear order.

Let us start with equation~(\ref{ma}). Linear order perturbations gives
\begin{align}
  \dot{\delta \rho_m} + 3(1+w)\frac{\rho^{(0)}_m}{a_0} \dot{\delta a} = 0 \,,
  \label{p1}
\end{align}
which can be integrated as
\begin{align}
  \delta \rho_m = -3(1+w)\frac{\rho^{(0)}_m}{a_0} \delta a \,,
  \label{p2}
\end{align}
which usefully connects the matter perturbation to the scale factor perturbation.

Now, we have three equations with three variables  $\delta a$, $\delta \phi$ and $\delta \rho_m$ subject to the equation (\ref{p2}) resulting from conservation
equation. So, we have just two independent equations to solve,
namely (\ref{fr2}) and (\ref{kg}).
We linearize equations (\ref{fr2}) and (\ref{kg}) using the perturbations and insert equation (\ref{p2}) and the background solutions (\ref{bg1})-(\ref{bg5}) in the field equations~(\ref{fr2}) and (\ref{kg}). The final results have been reduced to the following second order ordinary differential equations.
\begin{eqnarray}\label{eeq1}
  \ddot{\delta a}(t)-4\pi\rho^{(0)}_m(1+w)(1+3w)\delta a(t)=0,
  \end{eqnarray}
  \begin{eqnarray}
 \Big(\varepsilon-3\kappa\frac{K}{a^{2}_{0}}\Big)\ddot{\delta\phi}(t)+V''(\phi_{0})\delta\phi(t)=0.
  \label{p4}
\end{eqnarray}
The solution of equation (\ref{eeq1}) is given by
\begin{eqnarray}\label{solu}
\delta a(t)=d_{1}e^{\Omega t}+d_{2}e^{-\Omega t},
\end{eqnarray}
where $d_{1}$ and $d_{2}$ are the integration constants and $\Omega$ is
defined by
\begin{equation}
\Omega=\sqrt{4\pi\rho^{(0)}_m(1+w)(1+3w)}~~.
\end{equation}The solution is stable within the following ranges
\begin{equation}\label{int1}
w>-1/3,
\end{equation}
\begin{equation}\label{int}
-1<w<-1/3.
\end{equation}
Note that the interval (\ref{int}) violates the strong energy condition $\rho+3p\geq0$.
To complete our study we must consider stability of the scalar field equation (\ref{p4}).
Thus, we must consider stability of equation (\ref{p4}), for $K=0$, $K=1$ and $K=-1$.
We can rewrite equation (\ref{p4}) as follows

\begin{eqnarray}
\ddot{\delta\phi}(t)-\frac{V''(\phi_{0})}{3\kappa\frac{K}{a^{2}_{0}}-\varepsilon}\delta\phi(t)=0,
  \label{p44}
\end{eqnarray}
which provides us with the solution
\begin{eqnarray}\label{solu2}
 \delta \phi=C_{1}e^{\lambda t}+C_{2}e^{-\lambda t},
 \end{eqnarray}
where $C_{1}$ and $C_{2}$ are constants of integration, and $\lambda$ is
defined by
\begin{equation}
\lambda=\sqrt{\frac{V''(\phi_{0})}{3\kappa\frac{K}{a^{2}_{0}}-\varepsilon}}~.
\end{equation}
The stability condition of the scalar field for three following cases are given by
\begin{itemize}
\item{\it Flat Universe }($K=0$)
\begin{equation}\label{11}
\frac{V''(\phi_{0})}{\varepsilon}>0,
\end{equation}
\item{\it Closed Universe} $(K=1)$
\begin{equation}\label{22}
\frac{V''(\phi_{0})}{3\kappa\frac{1}{a^{2}_{0}}-\varepsilon}<0,
\end{equation}
\item{\it Open Universe} $(K=-1)$
\begin{equation}\label{33}
\frac{V''(\phi_{0})}{3\kappa\frac{1}{a^{2}_{0}}+\varepsilon}>0.
\end{equation}
\end{itemize}
To become more specific, we take  a typical choice for the potential $V(\phi)$ and examine the stable static solutions. For instance, consider the potential given by
\begin{align}
V(\phi) = \frac{1}{2} m^2 \phi^2 \,,
\end{align}
where $m$ is a positive constant scalar field mass. The potential satisfies $V''(\phi) = m^2$ and is strictly positive.
Thus, from equations (\ref{11}), (\ref{22}) and (\ref{33}) we get $\varepsilon>0$, $a^{2}_{0}>\frac{3\kappa}{\varepsilon}$ and $a^{2}_{0}>-\frac{3\kappa}{\varepsilon}$ for the cases $K=0$, $K=1$ and $K=-1$, respectively. It can be seen that in the case of $K=1$, the stability condition for the scalar field results in $a_0>\sqrt{\frac{3\kappa}{\varepsilon}}$. Hence, assuming $\kappa\ll{\varepsilon}$
accounts for an small initial size for the Einstein static Universe before the inflationary era. In particular, if we suppose that the non-minimal coupling is set to the Planck length, by quantum gravity and quantum cosmology considerations, as $\kappa=L_P^2$, then we find the inequality $a_0>\sqrt{3}L_P$ which asserts that the initial size of the Einstein static Universe must be greater than the Planck length.
\section{Vector and Tensor Perturbations}\label{sec:5}
In the cosmological background, the vector perturbations of a perfect fluid with equation of state, $p = w\rho$, are ruled by the co-moving dimensionless vorticity defined as $\varpi_a = a\varpi$. Thus, the vorticity
modes obey the following propagation equation\cite{5}
\begin{align}\label{vor}
\dot{\varpi}_{k}+(1-3c_{s}^{2})H\varpi_{k}=0 \,,
\end{align}
where $c^{2}_s = dp/d\rho$ is the sound speed and $k$ denotes the co-moving index (see below). This equation is valid in our consideration of Einstein static Universe in the framework of the non-minimal kinetic coupled gravity through the modified Friedmann equations (\ref{fr1}) and (\ref{fr2}). For the Einstein static background, Eq.(\ref{vor})
reduces to
\begin{align}
\dot{\varpi}_{k}=0 \,.
\end{align}
It can be seen that from the above equation, initial vector perturbations remain frozen and therefore we have neutral stability
against vector perturbations.

Gravitational-wave perturbations, namely tensor perturbations, of a perfect fluid is explained by the
co-moving dimensionless transverse-traceless shear $\Sigma_{ab}=a\sigma_{ab}$, whose modes satisfy the following equation
\begin{align}
\ddot{\Sigma}_{k}+3H\dot{\Sigma}_{k}+\Big[\frac{k^{2}}{a^{2}}+\frac{2K^{2}}{a^{2}}
-\frac{8\pi}{3}(1+3w)\rho+\frac{2}{3}\Lambda\Big]\Sigma_{k}=0 \,,
\end{align}
where use has been made of $D^{2} \rightarrow -k^{2}/a^{2}$ in which $D^{2}$ is the covariant spatial Laplacian. For the Einstein static background in our model, this equation becomes
\begin{align}
\ddot{\Sigma}_{k}+4\pi\rho_{0}(k^{2}+2K)(1+w)\Sigma_{k}=0 \,.
\end{align}
In order for the Einstein static Universe becomes stable against the tensor perturbations, and considering $\rho_0>0$, we have to require
\begin{equation}
(k^{2}+2K)(1+w)>0,
\end{equation}
which yields the following conditions
\begin{itemize}
\item $w >-1$ for $K=0, 1$\,,
\item $w >-1$ for $K=-1$ and $k^{2}>2$\,,
\item $w <-1$ for $K=-1$ and $k^{2}<2$\,.
\end{itemize}
The first two conditions show the equation of state parameter below the phantom divide, while the latter condition shows the equation of state parameter above the phantom divide. Considering the stability conditions (\ref{int1}) and (\ref{int}), it turns out that the first two conditions for stability against the tensor perturbations are consistent with the conditions for stability against the homogeneous scalar perturbations, except for $\omega=-1/3$. On the other hand, the latter condition for stability against the tensor perturbations is inconsistent with the stability conditions (\ref{int1}) and (\ref{int}) against the homogeneous scalar perturbations.
Therefore, in an open Universe with $k^{2}<2$, It is not possible to have
  stability against both scalar and tensor perturbations.

\section{Summary and Discussion}\label{sec:concl}

Non-minimal kinetic coupled gravity is one the novel modification to the general relativity, which includes the non-minimal coupling  of kinetic term of a scalar field $\phi$ with the curvature tensor by the coupling $\kappa$, and the minimal coupling with the metric tensor by the coupling $\varepsilon$. It has already been shown that such modified gravity model provides an essentially new inflationary mechanism. In this work, motivated by the idea of ``Emergent Universe'' of Ellis {\it et al}, we have assumed that the Universe might have been started out in an asymptotically Einstein static state as a initial state before the inflationary stage of the Universe. Then, we have studied the stability of  Einstein static Universe by using linear homogeneous perturbations in non-minimal kinetic coupled gravity. By taking a linear equation of state parameter for the matter distribution, the stability regions of the Einstein static Universe are specified by the second derivatives of the scalar potential. We have shown that the stability of Einstein static Universe, in the non-minimal kinetic coupled gravity with quadratic scalar field potential, for closed isotropic and homogeneous FLRW ($K=1$) Universe depends on the coupling parameters $\kappa$ and $\varepsilon$, such that in order to have an small initial size for the Einstein static Universe, consistent with the quantum gravity and quantum cosmology requirements (that the Planck length is the minimum possible length for the size of the Universe), the best choice is $\varepsilon=1$ and $\kappa=L_P^2$. Thus, the order of magnitude of the coupling parameter $\kappa$ has been determined in the present study of  Einstein static Universe
in the framework of kinetic coupled gravity.

From the cosmological point of view, the non-minimal kinetic coupled gravity is used for introducing new inflation models at early stage of the Universe. In order for this theory is considered as an alternative theory of general relativity, the contribution of the non minimal coupling should fade away for late times, so that both theories coincide with each other at low energy scale. The non minimal coupling includes the dominant term $\kappa G^{00} \sim \kappa H^2$. During the inflationary period we have $H\gg 1$, so that the role of $\kappa H^2$ is considerable in the cosmic dynamics. However, after the inflation, when the Hubble parameter is smaller, this term should be almost vanishing in order to recover the general relativity. Hence, it seems the coupling parameter plays an important role to justify this requirement. In this paper, we have found that assuming $\kappa=L_P^2$, the stability condition imposes the inequality $a_0>\sqrt{3}L_P$ on the initial size $a_0$ of the closed Einstein static Universe before the inflation. Such inequality asserts that the initial size of the Einstein static Universe must be greater than the Planck length $L_P$, in consistency with the quantum gravity and quantum cosmology requirements. Therefore, it seems a very small $\kappa=L_P^2$ not only provides us with a suitably small size Einstein static Universe,
but also is favored in kinetic coupled gravity to be considered as an alternative theory of general relativity.  This may be considered as a novel cosmological viable condition imposed on the kinetic coupled gravity theory.

We have considered a quadratic scalar field potential. If the other forms of $V(\phi)$  are taken,  the qualitative results will not change drastically.
Actually, the quadratic potential is a typical example of those potentials which satisfy the condition $V''(\phi)>0$
at the static point $\phi_0$. So, it is expected that by choosing other forms of $V(\phi)$ satisfying this condition, the qualitative results do not change drastically.

Finally, we have studied the stability against the vector and tensor perturbations. We have found the neutral stability against the vector perturbations, and discussed on those acceptable values of the equation of state parameter which yield stability against both homogeneous scalar perturbations and tensor perturbations.

\section*{Acknowledgments}
This work has been supported financially by a grant number 217/D/10856 from Azarbaijan Shahid Madani University.

\end{document}